\documentclass{aastex}
\usepackage{amsbsy,emulateapj5,epsfig,onecolfloat,apjfonts}   

\shorttitle{Constraints on the Energy Density From Galaxy Clusters}
\shortauthors{S.~M.~Molnar, Z.~Haiman, M.~Birkinshaw and R.~F.~Mushotzky}

\def\CHANDRA{{\it Chandra }}

\def\XMM{{\it XMM-Newton }}

\catcode`\@=11
\def\gsim{\ifmmode{\mathrel{\mathpalette\@versim>}}
    \else{$\mathrel{\mathpalette\@versim>}$}\fi}
\def\lsim{\ifmmode{\mathrel{\mathpalette\@versim<}}
    \else{$\mathrel{\mathpalette\@versim<}$}\fi}
\def\@versim#1#2{\lower 2.9truept \vbox{\baselineskip 0pt \lineskip 
    0.5truept \ialign{$\m@th#1\hfil##\hfil$\crcr#2\crcr\sim\crcr}}}
\catcode`\@=12

\begin{document}

\twocolumn
[
\title{Constraints on the Energy Content of the Universe \\
       From a Combination of Galaxy Cluster Observables}

\author{
Sandor M. Molnar\altaffilmark{1}, Zolt\'an Haiman\altaffilmark{2}, Mark Birkinshaw\altaffilmark{3} and
 Richard F. Mushotzky\altaffilmark{4}
}

\vspace{\baselineskip}

\begin{abstract}
We demonstrate that constraints on cosmological parameters from 
the distribution of clusters as a function of 
redshift ($dN/dz$) are complementary to accurate angular diameter distance 
($D_A$) measurements to clusters, and their combination significantly
tightens constraints on the energy density content of the Universe.
The number counts can be obtained from X--ray and/or SZ
(Sunyaev-Zel'dovich effect) surveys, and the angular diameter
distances can be determined from deep observations of the
intra-cluster gas using their thermal bremsstrahlung X--ray emission
and the SZ effect.  We combine constraints from
simulated cluster number counts expected from a 12 deg$^2$ SZ cluster
survey and constraints from simulated angular diameter distance
measurements based on the X-ray/SZ method assuming a statistical accuracy 
of 10$\%$ in the angular diameter distance determination of 100 clusters 
with redshifts less than 1.5.  We find that $\Omega_m$ can be determined
within about 25$\%$, $\Omega_\Lambda$ within 20$\%$, and $w$ within
16$\%$.  We show that combined $dN/dz + D_A$ constraints can be used
to constrain the different energy densities in the Universe even 
in the presence of a few percent redshift dependent systematic error in $D_A$.
We also address the question of how best to select clusters
of galaxies for accurate diameter distance determinations.  We show
that the joint $dN/dz + D_A$ constraints on cosmological parameters
for a fixed target accuracy in the energy density parameters are
optimized by selecting clusters with redshift upper cut--offs in the
range $0.5\, \lsim \;z\; \lsim \,1$.
\end{abstract}

]

\altaffiltext{1}{Department of Physics and Astronomy, Rutgers University, 136 Frelinghuysen Road, Piscataway, NJ~08854}

\altaffiltext{2}{Department of Astronomy, Columbia University, 550 West 120th Street, New York, NY 10027}

\altaffiltext{3}{Department of Physics, University of Bristol, Tyndall Avenue, Bristol, BS8 1TL, UK}

\altaffiltext{4}{NASA/Goddard Space Flight Center, Laboratory for High Energy Astrophysics, Greenbelt, MD 20771}

\section{Introduction}
\label{s:Introduction}

A point often raised in the era of ``precision cosmology'' is that any
particular method to determine fundamental cosmological parameters
will suffer from degeneracies: only a combination of parameters can be
determined accurately (Bridle et al. 2003). This issue has been
highlighted recently by a number of papers focusing on the
cosmological usefulness of galaxy clusters (Ettori, Tozzi \& Rosati 2003; 
Vikhlinin et al. 2003; Mei \& Bartlett 2003; Kujat et al. 2002; Newman et al. 
2002; Rubino-Martin \& Sunyaev 2002; Verde, Haiman \& Spergel 2002; 
for a recent review based on X-ray observations see Rosati, 
Borgani \& Norman 2002).  Indeed, to date the 
tightest constraints have been derived using a combination of
different methods, as demonstrated recently by the {\it Wilkinson
Microwave Anisotropy Probe (WMAP)} team (Spergel et al.  2003).  The
importance of using independent methods to determine cosmological
parameters is two--fold: (1) the methods may be combined to break
degeneracies and better constrain individual parameters; and (2)
consistency tests are possible and systematic errors can be studied.

Although we focus on constraints from the change in the number density 
of clusters as a function of redshift, $dN/dz$, 
and from measurements of the angular diameter distance of clusters based on 
the Sunyaev-Zel'dovich (SZ) effect, $D_A$,
further cluster observables could be used as additional indicators 
of cosmological parameters. Levine et al. (2002) showed that using the cluster
temperature function to constrain cosmological parameters can be made
more efficient by adding the mass-temperature relation normalization 
from X-ray observations, and discussed briefly how to
optimize cluster surveys based on their method.  Majumdar \& Mohr
(2003) emphasized that the mass-temperature relation with masses
determined in a relatively modest follow--up program can be used in
combination with the $dN/dz$ test to greatly improve systematic
limitations that arise from cluster evolution.  Several other
observables can be useful, such as scaling relations (Bialek et
al. 2001; Verde et al. 2002), the shape of the cluster mass function
(Hu 2003); and the three--dimensional cluster power spectrum (Refregier
et al. 2002).

Constraints on $w$ as a function of redshift from planned SZ and supernovae 
surveys, and from their combination were studied by Weller, Battye \& Kneissl 
(2002) and Weller \& Albrecht (2002). Note, however, that cosmic variance can 
also be important in some cluster surveys (Hu \& Kravtsov 2003).

The redshift distribution of the number of clusters, $dN/dz$, is sensitive to
both the change of the cosmological volume element and to the growth
function of structure formation.
Cluster number counts as a function of redshift can be determined from
X--ray or SZ surveys (e.g. Carlstrom et al. 2002; Rosati et al. 2002).
Haiman, Mohr \& Holder (2001)
discussed in detail constraints on dark energy (via the equation of state parameter, $w$) 
from the redshift distribution of clusters
from future SZ and X--ray surveys and suggested that
combining those constraints with constraints using Type Ia SNe or CMB
fluctuations can lift degeneracies in the $\Omega_m$-$w$ plane
(see also Wang \& Steinhardt 1998). Holder, Haiman \& Mohr (2001) used 
the redshift distribution of clusters
from future SZ and X--ray surveys to constrain $\Omega_m$
and $\Omega_\Lambda$ pointing out that, similarly to the parameter space
($\Omega_m$,$w$), constraints from cluster redshift distribution are
complementary to constraints from Type Ia SNe and CMB fluctuations.
The X--ray thermal bremsstrahlung emission and SZ effect (Sunyaev \& Zel'dovich 1980) 
depend on different combinations of the physical
parameters of the cluster and cosmology and provide us with a way to
determine the angular diameter distance, $D_A$, to the cluster 
(the so--called X-ray/SZ method, for recent, detailed references, 
see Carlstrom, Holder \& Reese 2002 and Reese et al. 2003).  
The angular diameter distance probes directly the curvature of the Universe.

Molnar, Birkinshaw \& Mushotzky (2002) discussed constraints on models with 
parameters ($\Omega_m$,$w$,$h$), and ($\Omega_m$,$\Omega_\Lambda$,$h$) using 
simulations of angular diameter distance measurements to clusters of
galaxies and showed that the degeneracies in cosmological
parameters from this technique are similar to those from SNe Ia, and therefore
they are complementary to constraints from redshift distribution of clusters.
This is not surprising, since the luminosity distance, which is utilized in the 
SN studies, is closely related to the angular diameter distance.

In this paper, we show that it will be possible using clusters
of galaxies alone to constrain accurately the matter density parameter, 
$\Omega_m$, and the 
cosmological constant density parameter, $\Omega_\Lambda$, or the
equation-of-state parameter, $w = p/\rho$, by combining $dN/dz$ and
$D_A$ measurements of clusters using the X-ray/SZ method, 
and estimate the accuracy achievable in these parameters.

Since long observations are needed for high accuracy distance measurements 
with present day observatories, a relatively small sample of
clusters has to be selected.  This cluster sample does not have to be
assembled from the same survey that is used for the cluster redshift 
distribution test. 
Any sufficiently accurate survey would suffice as long as it has the 
necessary redshift coverage.  
 We address the natural question of how best to 
 select samples of clusters of galaxies for efficient angular
 diameter distance determination.
As an example, we discuss how we can optimize the selection of clusters for
detailed distance measurements by allowing both the number of clusters
in the sample and the upper redshift cut--off of the sample to be free
parameters.
We also discuss how to select a cluster sample to minimize the random errors.

The rest of this paper is organized as follows. In \S~2, we
briefly summarize our methodology. In \S~3, we apply our technique to
future cluster samples, and derive constraints on cosmological
parameters. In \S~4, we discuss the issue of the optimal selection of
clusters for angular diameter distance measures. Finally, in \S~5, we
offer our conclusions and summarize the implications of this work.

\section{Methodology}
\label{s:Methodology}

We have studied combined constraints from the redshift distribution of clusters 
and angular diameter distance measurements in two different
sets of models. Both sets are described by the usual parameters
($\Omega_m$,$\Omega_\Lambda$,$w$,$h$,$\sigma_8$,$n$), and we choose as
our fiducial model
($\Omega_m=0.3$,$\Omega_\Lambda=0.7$,$w=-1$,$h=0.65$,$\sigma_8=0.9$,$n=1$).
In Model A, we assume a cosmological constant ($w=-1$), but allow
($\Omega_m$,$\Omega_\Lambda$) to vary independently.  In Model B, we
assume the universe is spatially flat ($\Omega_m+\Omega_\Lambda=1$),
but allow $w$ to vary.  We used constraints from previous simulations
based on the redshift distribution of clusters derived
from a deep 12 deg$^2$ SZ survey (see Holder et al. 2001,
Model A; and Haiman et al. 2001, Model B). We then
determined the expected constraints on the same parameters from
$D_A$ measurements with fixed accuracy, changing the total number of
clusters and the upper redshift cutoff of the sample (for a
detailed description of the method see Molnar et al. 2002).

Most models of the X--ray emission from clusters of galaxies are based
on the assumption of hydrostatic equilibrium. Note, however, that this is not
a necessary assumption for the X-ray/SZ method to work.
Detailed, spatially resolved X--ray spectroscopy can be used to observe the 
projected temperature and X-ray surface brightness distribution
of the cluster and derive the de-projected temperature and gas density 
distribution, which, combined with the assumption of hydrostatic 
equilibrium, can be used to determine the total mass distribution of the
cluster (see for example Evrard et al. 2002).  
Recent XMM imaging spectra (Pratt \& Arnaud 2003) support previous 
observations of clusters which show that outside the central regions 
a large fraction of clusters are roughly isothermal and that the surface
density can be described by a $\beta$ model (Jones \& Forman 1984;
Mohr, Mathiesen \& Evrard 1999), where the gas density can be written
as $\rho(r)=(1+(r/r_c)^2)^{-3\beta/2}$.  The core radius, $r_c$, sets 
the scale of the cluster atmosphere and $\beta$ specifies the slope of
the density profile at large radius (Cavaliere \& Fusco-Femiano 1976).
For the sake of concreteness, we here assume an isothermal $\beta$
model for the intra-cluster gas, however this assumption is not
essential in the determination of $D_A$. 
The parameters that we need are: 
(1) $r_c$ and $\beta$, which define the spatial structure of the
    intra-cluster gas and can be deduced from X-ray or SZ images; 
(2) the amplitude of the X-ray brightness and SZ decrement/increment 
    (depending on the observing wavelength);
(3) the temperature of the intra-cluster gas and the X--ray absorbing
    column density; and 
(4) the redshift of the cluster 
    (derivable from x-ray or optical observations).

The most important known redshift independent systematic errors in
determining angular diameter distances arise from
calibration errors for the SZ and X-ray instruments, and finite cluster 
sizes from modeling the intra-cluster gas. Calibration errors cause
systematic scale errors in the peak intensities of the SZ effect
and X-ray surface brightness. The calibration errors of about
2.5$\%$ for interferometric observations (Carlstrom et al.~2002),
and about 10$\%$ for {\it XMM-Newton\/} and {\it Chandra},
induce 5$\%$ and 10$\%$ errors in $D_A$.
The $\beta$ model may give a divergent cluster mass if not truncated at some 
finite upper cut off radius. Assuming a $\beta$ 
model with infinite extent introduces a systematic underestimate of $D_A$.
Based on numerical simulations, Inagaki et al. (1995) estimated 
this systematic error to be as large as 10-20$\%$. Systematic error estimates based
on observations are much lower (up to 6$\%$, Holzapfel et al. 1997; 
Birkinshaw \& Hughes 1994). However, in this paper
we focus on the energy density and equation of state parameters which
are not sensitive to redshift independent systematic errors since they
depend only on the shape of the angular diameter distance-redshift
function (for a detailed analysis see Molnar et al. 2002).

Systematic errors depending on redshift do change the shape of the
$D_A$-$z$ relation, and therefore could bias the inferred energy
density parameters.  The most important such candidate is small scale
gas clumping that varies with redshift. A general discussion 
of the effect of clumping on the determination of angular diameter distance
can be found in Reese et al.~(2003) and Molnar et al.~(2002), and references therein.
Numerical simulations show that clumping caused by accretion shocks and mergers
results in an overestimate in the angular diameter distance to clusters
by about 30$\%$ (LCDM, Mathiesen, Evrard \& Mohr 1999). 
The error from clumping based on observations is estimated to be about 20$\%$
(Reese 2003).
We will ignore this possibility here and simply note that 
(1) at present, there is no observational evidence for redshift dependent 
    clumping, and 
(2) it is likely that a redshift dependent systematic effect would 
    cause different systematic errors in $dN/dz$ and $D_A$ measurements, 
    and thus these kind of errors could be identified based on consistency. 
Overall a 30$\%$ error in the angular diameter distance is well within reach
for a large sample of future surveys ($\ge$ 1000) using the spherical isothermal $\beta$
model (Reese 2003). We expect a much better accuracy for a smaller sample of clusters 
using spatially resolved spectroscopy.

 We assume further that the main known random error in determining $D_A$, the
  orientation bias for non-spherical clusters, has been solved for our
  sample using galaxy velocity dispersion measurements, or a
  combination of X-ray, SZ, and weak lensing data with the assumption
  of hydrostatic equilibrium (Fox \& Pen 2002).
While this is an optimistic assumption, we find that in a sample of 100 clusters
this effect would inflate our error estimates for $\Omega_m$ and $\Omega_\Lambda$ 
by only an additional $\sigma = 5\%$. 
 We expect about an 5$\%$ error in the cluster parameters from uncertainty in the
 spatial fitting, and a 5$\%$ error in the X-ray temperature is
 achievable with {\it XMM-Newton\/} and {\it Chandra} (Pratt \& Arnaud 2003). 
For a detailed analysis of the error budget of the X-ray/SZ method, see
Molnar et al. (2002), a summary of errors based on observations can be found in 
Reese (2003).
 We assume an additional 5$\%$ statistical error from 
 other sources.  Finally we assume that the Hubble constant is determined from other 
 measurements.

%
%

\begin{figure}[t]
\centerline{
\plotone{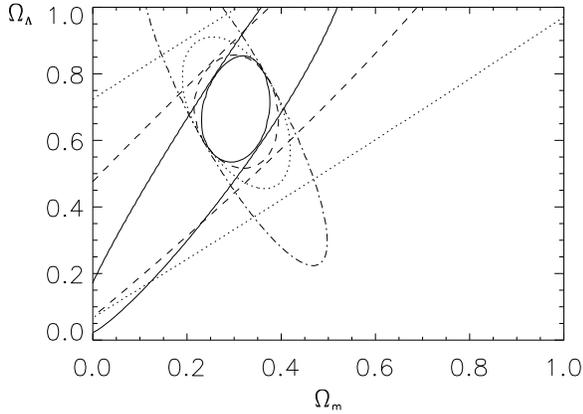}
}
\caption{Simulated 97\%
     confidence limits in the $\Omega_m - \Omega_\Lambda$ plane based
     on $dN/dz$ and $D_A$ measurements. The outer set of solid,
     dashed, and dotted lines are constraints from the $D_A$ values
     alone for samples of clusters with upper redshift cutoffs of $z =
     1.5$, $1.0$, and $0.5$. The dot-dashed line is the constraint
     from the $dN/dz$ measured in a $12 \ \rm deg^2$ SZ survey
     alone. The corresponding combined confidence intervals are shown
     as the inner solid, dashed, and dotted curves.
\label{F:FIG1}
}
\end{figure}

%
%

\begin{figure}[t]
\centerline{
\plotone{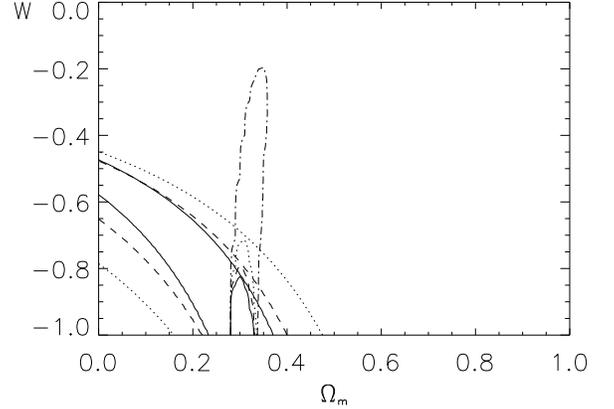}
}
\caption{Simulated 97$\%$ confidence limits in the $\Omega_m - w$ plane
     based on $dN/dz$ and $D_A$ measurements. 
     The line codes are the same as in Figure 1. 
\label{F:FIG2}
}
\end{figure}

Fox \& Pen (2002), using numerical simulations, demonstrated that 
 distances to clusters can be determined with a random error of about
 5$\%$ combining X-ray, SZ, and lensing measurements and the assumption of 
 hydrostatic equilibrium.  Although they ignored errors due to lensing, 
 their result suggest that a 10$\%$ error in the determination of angular 
 diameter  distances is realistic in the near future. Therefore we 
 performed simulations using a 10$\%$ error in the $D_A$ values.
 Note, however, that larger errors, as long as they are truly random, 
 can be compensated with using more clusters, for example, assuming a 20$\%$ 
 random error in $D_A$, we would need a  sample of about 400 clusters, and 
 our final conclusion of this paper would be unaltered. 
In all of our simulations the redshift distribution was derived by 
randomly sampling a uniform distribution in redshift space with an upper 
cut--off at fixed redshift. 
 The number of sampled clusters in the redshift
 range 0.1-1.5 changes by less than a factor of 3 as a function 
 of cosmology, which, in practice, makes it easy to choose a 
 quasi--uniform sample.  Note, however, that our conclusions are not 
 sensitive to the exact distribution of the number of clusters in redshift 
 as long as they smoothly cover the targeted redshift range.

Systematic effects in the cluster redshift distribution 
 from the mass--X-ray temperature, the power spectrum normalization, $\sigma_8$, 
 and the mass function were studied by Battye \& Weller (2003).
 They concluded that these systematic effects are important and 
 more studies are necessary to reduce them.
 Systematic effects in the cluster redshift distribution in SZ surveys
 from the surface brightness bias and 
 cluster spatial orientation have not been discussed in detail.
 The detectibility is a function of signal/noise and, for extended sources 
 (like clusters), this is determined by surface brightness and not by flux.
X-ray and SZ surveys will not detect very extended clusters which have surface 
brightness too low to yield the necessary signal/noise. 
 Even if we use a mass-size relation and
 a constant mass cut off, there is a dispersion around any observed or theoretical 
 mass-size relation.

As a result of the orientation bias, prolate clusters will be over-represented since 
they have larger central SZ and X-ray surface brightnesses than oblate clusters. 
 Faint ellipsoidal clusters aligned with their long axis in the plane of the
 sky might not be detected, lower mass clusters with their long axis in the
 line of sight will be falsely detected in a given mass bin. 
 However, there are more low mass than high mass clusters,
 so we underestimate the number of clusters at the given mass threshold.
 Using a mass cut-off which is higher than the nominal detection-limit-implied
 mass cut-off would give us a chance to estimate this bias.

We used the likelihood function obtained by Holder et al. (2001)
for model parameters $\Omega_m$ and $\Omega_\Lambda$ (our Model A).
Holder et al. defined their likelihood function based on the Cash $C$
statistic (Cash 1979), kept $h$ constant,  
and marginalized over the power spectrum normalization, $\sigma_8$.
We used the likelihood function of Haiman et al. (2001)
for the cluster redshift distribution test in 
parameter space $\Omega_m$ and $w$ (our Model B).
Haiman et al. defined their likelihood function as the
product of the Poisson probability of detecting a total number of $N$
clusters and a Kolmogorov-Smirnov probability for the unnormalized
redshift distribution of these $N$ clusters\footnote{Note that using
the likelihood function based on the Cash statistic would result in
somewhat tighter constraints (as demonstrated explicitly by a
comparison of these two different statistics in Verde et al. 2001).}.
Haiman et al. also kept $h$ constant and marginalized over the power
spectrum normalization.  
We kept $h$ constant, assumed Gaussian errors in the angular
diameter distance measurements, and constructed the likelihood function 
based on the usual $\Delta\chi^2$ statistic.
The angular diameter distance function does not depend on
the power spectrum, so we can combine these two likelihood functions
in a straightforward way by multiplying them. 
In more detailed work on specific observational strategies it would be
necessary to multiply the likelihood functions before
marginalizing over unconstrained parameters.

\section{Combined Constraints on Cosmological Parameters}
\label{s:Results}

In our first set of simulations, we assumed 100 clusters with an error
of 10$\%$ in the $D_A$ determinations with different redshift cut--offs
out to $z=1.5$.  We show the 97$\%$ confidence limits (CLs) for the
combined constraints from the redshift distribution of clusters
and angular diameter
distance measurements based on simulations in Figures 1 and 2.  In
Figure 1 we show constraints on $\Omega_m$ and $\Omega_\Lambda$ (Model
A). This figure shows that, as we select
clusters with higher and higher redshift cut--offs, the constraints
from $D_A$ measurements alone get tighter and the confidence contours 
rotate counter--clockwise.  The constraints get
tighter because at higher redshifts, $D_A(z)$ is more sensitive to
these parameters, and rotate, since these parameters are sensitive to
different combinations of the density parameters at different
redshifts.  As a consequence of this rotation, although at higher
upper redshift cut--offs the constraints from $D_A$ measurements 
are tighter, they are farther from orthogonal to constraints from the 
redshift distribution of clusters. The combined
constraints do not improve substantially from upper cut--offs of about
$z\gsim 1$, although the likelihood function does get somewhat narrower.

%
%

\begin{figure}[t]
\centerline{
\plotone{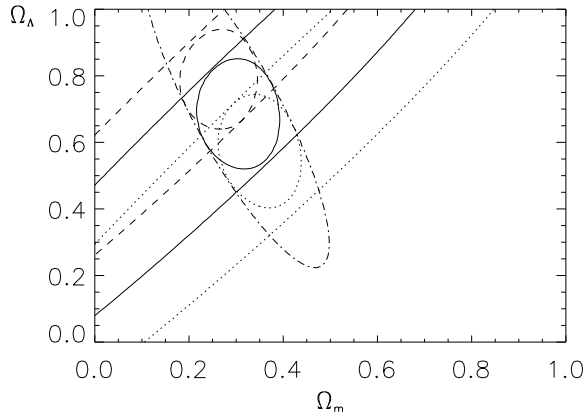}
}
\caption{Simulated 97\%
     confidence limits in the $\Omega_m - \Omega_\Lambda$ plane based
     on $dN/dz$ and $D_A$ measurements with redshift dependent systematic errors.
     The outer set of solid, dashed, and dotted lines are constraints from the 
     $D_A$ values alone for a sample of 100 clusters with random errors of 10$\%$, 
     and redshift dependent systematic errors of +5$\%$ and -5$\%$ at z=1 growing 
     linearly from 0 at z=0. The dot-dashed line is the constraint from the $dN/dz$ 
     measured in a $12\ \rm deg^2$ SZ survey alone. 
     The corresponding combined confidence intervals are shown as the inner solid, 
     dashed, and dotted curves.
\label{F:FIG1}
}
\end{figure}

%
%

\begin{figure}[t]
\centerline{
\plotone{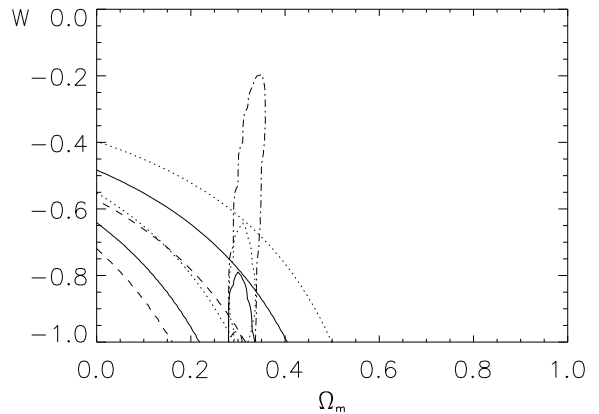}
}
\caption{Simulated 97$\%$ confidence limits in the $\Omega_m - w$ plane
     based on $dN/dz$ and $D_A$ measurements with redshift dependent systematic errors.
     The line codes are the same as in Figure 3. 
\label{F:FIG2}
}
\end{figure}

We conclude that, contrary to naive expectations, observations of
clusters at redshifts exceeding $z\approx 1$ will not improve the
constraints on $\Omega_m$ and $\Omega_\Lambda$ using these
methods. Rather, the most important parameter is simply the total
number of clusters, as long as we observe clusters with redshifts up
to $z\sim 1$.  This is good news because, while it is difficult to
determine distances at high redshifts ($z\gsim 1$) due to the low flux
of these clusters in the X--ray band, recent \XMM and \CHANDRA
observations readily find clusters out to
redshifts of about $z=0.9$ (see for example Ettori et al. 2002).  
Based on our simulations we find that the combined constraints from 100
clusters uniformly distributed with an upper cut--off at $z=1$ could be 
used to determine $\Omega_m$ to within about 25$\%$, and $\Omega_\Lambda$ 
to within 20$\%$ (97$\%$ CL).

We show our results on combined constraints on $\Omega_m$ and $w$
(Model B) from simulations in Figure 2.
The figure shows that the constraints from angular diameter
distances alone are once again getting tighter as we choose higher
redshift cut--offs, but the constraints also spread more in $w$ 
(they cross the $w$ axis at higher values of $w$). 
As a result, as it can be seen from this figure, the high-$w$ limit does 
not improve by choosing clusters with redshift cut--offs greater
than about $z\approx 1$. Using 100 clusters, the combined method could
determine $w$ to about 16$\%$ (97$\%$ CL).

 We choose to study constraints on the energy densities in the Universe
 because they are not sensitive to systematic effects which are redshift
 independent. However, the energy densities are sensitive to redshift 
 dependent systematics. 
 Although such an effect has not yet been found, a few percent redshift 
 dependent systematic error would be difficult to identify observationally. 
 We carried out simulations to study how constraints 
 on $\Omega_m$ and $\Omega_\Lambda$ (Model A) and on $\Omega_m$ and $w$
 (Model B) change assuming redshift dependent systematics at few percent 
 level. 
 We show our results on Figures 3 and 4 assuming a systematic bias in the 
 inferred distance that grows linearly from no bias at z=0 to a +(-)5$\%$ 
 over(under)estimate of all distances at z=1 using 100 clusters and
  10$\%$ random errors, as before. Combining the $D_A$ and $dN/dz$
  results still permits the measurement of the densities of matter and
  dark energy, or the density of matter and the equation of state
  parameter. Although a systematic shift in the derived values of
  these parameters is caused, such 5$\%$ systematic errors do not move
  the error ellipse away from the input model values significantly. 
  While larger redshift-dependent systematic errors of this type
  would lead to significant systematic errors in the derived
  parameters, they would also cause inconsistencies in the $D_A$ and $dN/dz$
  tests through which they could be detected.

\section{Optimizing the Sample Selection for Distance Measurements}

Although we expect hundreds of clusters to be discovered in future X--ray 
and SZ surveys (for example, the \XMM 64 deg$^2$ Large Scale Structure 
Survey will provide hundreds of candidate clusters, Refregier et al. 2002),
it is not feasible to determine accurate distances to all discovered
clusters.  The reason for this is two--fold: (1) the observational
time necessary for accurate distance determination rapidly increases
with distance, and is on the order of 50-100 ks with \XMM and
\CHANDRA at redshifts of about 0.7; and (2) not all clusters are relaxed.

 The question then naturally arises: How can we assemble our cluster
 sample optimally, yielding the most precise and robust
 constraints on cosmological parameters?  A fundamental physical
 selection criterion is that the clusters should be as relaxed as possible. 
 Based on numerical simulations, Roettiger, Stone \& Mushotzky (1997), 
 in order to minimize the effects of dynamical activity, 
 suggest to exclude clusters with non-cylindrical surface brightness
 and temperature distribution in projection, with twisted isophots, 
 with anisotropic galaxy velocity distribution, and those clusters 
 for which $\beta$ from spatial fitting is significantly different from $\beta$ 
 determined from spectroscopy.
 If we wish to eliminate the orientation bias, we need to use clusters
 which are, in most part, in hydrostatic equilibrium. 
 Although the X-ray/SZ method works for any 
 cluster as long as we have an accurate physical model for the gas,
 clusters which are not in dynamical equilibrium are difficult to model, 
 contribute a large systematic error, and so should be excluded. 
 The signature of such objects is complicated morphology of the 
 spatial structure in the X--ray surface brightness and/or
 temperature image. 
 Line of sight mergers would be difficult to identify based on morphology. 
 Mergers might be identified by comparing distances determined using
 the X-ray/SZ and other methods. 
 The velocity distribution of member galaxies can also be used to 
 identify line of sight merging and numerical simulations can be 
 used to verify merging (Gomez, Hughes \& Birkinshaw 2000).
  A cluster which is an outlier in scaling relations (X-ray temperature -- luminosity,  
  mass -- X-ray temperature, mass -- X-ray luminosity)
  would be a likely case of dynamical activity, since on-going merging
  leads to temporary enhancements of the X-ray flux and raised average intra-cluster 
  gas temperatures because of shock heating (Randall, Sarazin \& Ricker 2002).

%
%

\begin{figure}[t]
\centerline{
\plotone{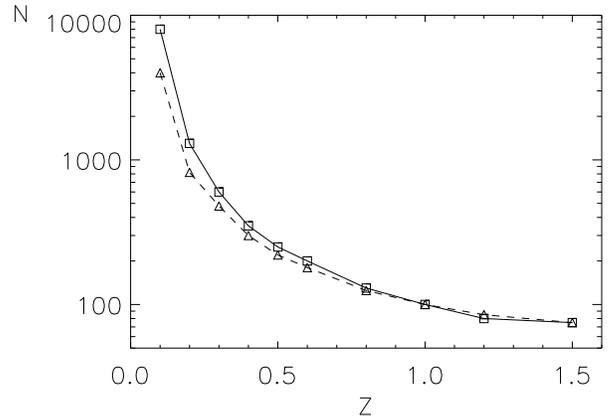}
}
\caption{The number of clusters, $N$, in the sample for angular diameter
distance determination that is necessary to achieve a fixed
statistical accuracy, 25$\%$ in $\Omega_m$, 20$\%$ in $\Omega_\Lambda$
(Model A) and 16$\%$ in $w$ (Model B) from combining SZ survey and $D_A$ 
measurements, as a function of the upper redshift cut--off of the $D_A$ sample. 
The squares and the solid curve represent the number of clusters necessary
to determine $\Omega_m$ and $\Omega_\Lambda$, the triangles and the
dashed curve represent the number of clusters needed to determine $w$.
\label{F:FIG3}
}
\end{figure}

In our previous simulations, we kept the number of clusters fixed at
$N=100$, in order to see whether constraints improve by using clusters
at $z > 1$. Since it takes longer to achieve 
the target accuracy at higher redshifts, one has to consider the
observational time as well.  A detailed analysis of how the required
observational time for a given $D_A$ accuracy depends on the redshift
of the cluster is beyond the scope of this paper since the
necessary parameters are instrument specific.  Instead we invert
the question, and use simulations to determine the
number of clusters necessary to observe to achieve a fixed accuracy on
the cosmological parameters, as a function of the upper redshift
cut--off for the $D_A$ sample. For concreteness, we fix the errors at
the level achievable with observations of 100 clusters with 10$\%$
accuracy in the angular diameter distance determination with an upper
cut--off of $z=1$, i.e. we determine the number of clusters 
necessary to achieve an accuracy of about 25$\%$ in
$\Omega_m$, and 20$\%$ in $\Omega_\Lambda$ (for our Model A), and
16$\%$ in $w$ (for our Model B) as a function of upper redshift cut-off 
in the $D_A$ cluster sample. 

In general, different numbers of clusters are necessary to achieve any
specified accuracy for different parameters. In practice, we find that the
ratio of errors does not change significantly in the considered
redshift range and therefore once the ratio between 
errors in $\Omega_m$ and $\Omega_\Lambda$ is fixed, it is possible to
determine the number of clusters in the sample to achieve the same
ratio in accuracy in both parameters as a function of redshift 
cut--off.  Since adding constraints from $D_A$ measurements to those from
the redshift distribution of clusters
in Model B does not improve constraints on the matter
density, we determine only the number of clusters necessary to achieve
a fixed accuracy in $w$ as a function of redshift.  We show our
results in Figure 5.  Interestingly, the curves give rather similar 
requirements for the number of clusters.  The figure also clearly shows that, 
as expected, with low upper cut--offs on redshifts of the sample, the
necessary number of clusters increases, since the angular diameter
distance is insensitive to the energy
  content of the Universe at low redshift. In accordance with our
  conclusions from Figures~1 and~2, Figure~5 shows that at redshift
  cutoffs $z\gsim 1$ the number of clusters needed tends to a fixed value.
However, the observing time will increase significantly if one wants to 
maintain the assumed 10$\%$ accuracy in the angular diameter distance 
measurements.  Overall, Figure~5 suggests that the optimal redshift range 
of a cluster sample for useful $D_A$ measurements is $0.5\lsim z \lsim 1$.

\section{Conclusions}

We have demonstrated that the energy content of the 
Universe can be constrained to high statistical accuracy using the 
cluster redshift distribution and angular diameter distance measurements.
Degeneracies in cosmological parameters that are constrained by either 
observable are substantially weakened when they are used in tandem.  
We quantified constraints from the simulated cluster redshift distribution 
expected from a 12 deg$^2$ SZ cluster survey
and constraints from simulated angular diameter distance measurements
based on using the X-ray/SZ method, assuming an expected accuracy of 10$\%$
in the angular diameter distance determination of 100 clusters with
redshifts $z\lsim 1.5$. We find that $\Omega_m$ can be determined to
a statistical accuracy (97$\%$ CL) of about 25$\%$, $\Omega_\Lambda$
within 20$\%$, and $w$ to an accuracy of about 16$\%$.  We also
addressed the question of how to select clusters of galaxies optimally for
accurate diameter distance determinations. Our results indicate that
the joint $dN/dz + D_A$ constraints on cosmological parameters for a
given observation time are optimized by using cluster $D_A$
measurements in the redshift range $0.5\lsim z \lsim 1$.
We carried out simulations to study how combined constraints on
the energy densities depend on redshift dependent systematic errors.
We found that a combination of cluster redshift distribution and 
angular diameter distance determination measurements with an assumed 
redshift dependent systematic error which grows linearly to 5$\%$ at 
$z=1$ still leads to significantly improved constraints on cosmological
  parameters, without unacceptable systematic errors in those parameters.

Comparison of the errors derived from the $dN/dz + D_A$ technique with
100 clusters compares favorably with those from simulations based on
type Ia supernovae (e.g. Gerke \& Efstathiou 2002; see also Hannestad
\& Mortsell 2002; Perlmutter et al. 1999), with similar errors on $w$,
but significantly tighter errors on $\Omega_m$.  Inclusion of priors
from the {\it WMAP} experiment will tighten confidence limits further
(complementarity between cosmic microwave background and cluster
$dN/dz$ constraints have been explicitly demonstrated in Haiman et
al. 2001).  We believe that the results in this paper are
important, because they indicate that independent, tight statistical
constraints on cosmological parameters will be available internally
from galaxy cluster surveys.  Having independent constraints from
different methods will be invaluable in understanding the systematic
errors that will likely limit our determination of cosmological
parameters in the next several years.

\acknowledgments

We thank Rashid Sunyaev for advice and encouragement, Joe Mohr for valuable 
comments on the manuscript, and the anonymous referee for constructive comments 
that allowed us to improve our paper. 
This work (SMM) was partly supported by NASA LTSA Grant NAG5-3432 
(PI: J. P. Hughes).

%
%


\end{document}